# Nanoscale local modification of PMMA refractive index by tip-enhanced femtosecond pulsed laser irradiation

Denis E. Tranca[1], Stefan G. Stanciu[1], Radu Hristu[1], Adrian M. Ionescu[2], George A. Stanciu[1]*

[1]University Politehnica of Bucharest, Romania
[2]École polytechnique fédérale de Lausanne, Switzerland
*Corresponding author: stanciu@physics.pub.ro

**Abstract**

Investigation techniques based on tip-enhanced optical effects, capable to yield spatial resolutions down to nanometers level, have enabled a wide palette of important discoveries over the past twenty years. Recently, their underlying optical setups are beginning to emerge as useful tools to modify and manipulate matter with nanoscale spatial resolution. We try to contribute to these efforts by reporting a method that we found viable to modify the surface refractive index of polymethyl methacrylate (PMMA), an acrylic polymer material. The changes in the refractive index are accomplished by focusing a femtosecond pulsed near-infrared laser beam on the apex of a metalized nano-sized tip, traditionally used in scanning probe microscopy (SPM) applications. The adopted illumination strategy yields circular-shaped modifications of the refractive index occurring at the surface of the PMMA sample, exhibiting a lateral size <200 nm, under 790 nm illumination, representing a four-fold increase in precision compared to the current state-of-the-art. The light intensity enhancement effects taking place at the tip apex makes possible achieving refractive index changes at low laser pulse energies (<0.5 nJ), which represents two orders of magnitude advantage over the current state-of-the art. The presented nanoimprinting method is very flexible, as it can be used with different power levels and can potentially be operated with other materials. Besides enabling modifications of the refractive index with high lateral resolution, this method can pave the way towards other important applications such the fabrication of photonic crystal lattices or surface waveguides.

**Keywords:**

Femtosecond laser nanoimprinting, refractive index change, tip-enhanced optical effects, waveguides.

**Introduction**

The possibility to precisely manipulate the refractive index (*n*) of a material represents a cornerstone for many important industrial optical applications [1–5] and the utility of related methods has also been demonstrated in biomedicine [6,7]. The requirement of novel methods yielding better precision, easier implementation, and lower costs is motivated by the continuously increasing attention and demand for light guiding applications [5,8], photonic crystals [9], and optoelectronic devices [10,11] with various properties and levels of performance. Furthermore, light manipulation at small scale (in terms of guiding and phase changing) together with bandgaps design in photonic crystals are generally acknowledged as key actors in the growing field of optical computation [12,13]. These are usually achieved by the mixed use of different materials exhibiting distinct refractive indices [14]. Other ways to achieve a desired distribution or pattern of refractive indices rely on photo-polymerization [15] and photolithography [16] approaches, in the frame of which the refractive index of a material is locally changed. A promising way to modify the local refractive index of some special glass types and polymers is to use a femtosecond (*fs*) laser beam focused on the surface of such samples [17–19]. This technique was successfully demonstrated on BK7 glass [20], borosilicate glass [21] and polymers [11,22–25]. However, such approaches lack lateral precision, with the size of the smallest sample areas for which the refractive index is modified being comparable to the beam's wavelength or even larger [22,24,26,27]. Furthermore, previous studies [28,29] discussing such advances report the use of laser pulse energies in the range of tens and hundreds of nJ, which are suitable only for materials that are not prone to thermal damage.

With respect to the physico-chemical effects responsible for the refractive index changes occurring in a certain material upon exposure to a fs laser beam, these depend on the material type. For polymers, past studies suggest that the nonlinear absorption occurring in the illuminated material is responsible for the refractive index changes that were observed with fs beam exposure [30]. Additionally, it was reported that the modified region exposed to a pulsed fs laser beam operating at a kilohertz repetition rate mode was directly related to the plasma density originating from the beam's intensity distribution [30]. Another reported cause for the same effect is the avalanche ionization [29,31] observed in PMMA (via continuous PMMA chains breaking) occurring in the presence of the electromagnetic field. This leads to increased electron density and high optical absorption which causes irreversible modifications of the material. Other possible cause of the refractive index change occurring in polymers is the depolymerization effect [32,33], which may occur when a polymer is exposed to fs laser beams, resulting in a decrease of the refractive index [32]. Physically, depolymerization is accompanied by volume expansion [32,34] and decrease of the Young's modulus [35,36]. Other relevant works carried thorough discussions on the mechanisms of refractive index changes occurring in PMMA upon fs beam irradiation have been presented by A. Baum et al. [32,33].

In the field of optical microscopy techniques capable of sub-diffraction limit resolution, apertureless scanning near-field optical microscopy (a-SNOM) has attracted great interest given its capabilities to provide lateral resolutions down to the nanometers level, under ambient conditions and in a label-free manner, while also providing chemical specificity [37,38]. One of the most prominent a-SNOM techniques is scattering SNOM (s-SNOM), which is based on an atomic force microscope (AFM) probe that scans the sample surface while being illuminated by a laser beam focused on its apex. The light scattered by the probe's tip harbors information about the optical properties of the sample, and besides other uses [38], s-SNOM images and spectra can be used to extract the complex refractive index of materials [39–43]. In the scattering process, the electric field enhancement, which occurs at the tip of the probe, plays a crucial role [44,45]. Several other imaging techniques that rely on tip-enhanced effects can be implemented with architectures that resemble in great detail an s-SNOM setup [46–52]. Imaging modalities such as tip-enhanced Raman, tip-enhanced fluorescence, tip-enhanced second harmonic generation or tip-enhanced photoluminescence yield complementary information at the nanoscale and represent valuable nano-characterization tools, enabling a thorough understanding of a sample's optical, chemical, and morpho-structural properties [46–52].

A series of past efforts have demonstrated the possibilities to use optical setups for tip-enhanced microscopy for the fabrication of nanostructures [53–55]. Part of these efforts involves the use of a fs laser source working in tandem with s-SNOM [53], aperture-SNOM [54] or other techniques (micro-lens array [56], laser-interference lithography [57], etc.). However, none of these previous efforts have focused, nor documented, refractive index changes occurring upon the laser beam enhancement by the tip. Furthermore, most of these past works use complex experimental setups or special materials (like porous glass immersed in water with fs laser exposure followed by annealing [58]) to generate different physical effects, which conduct to nano-structural modifications. In contrast to such special designed materials, PMMA is a common acrylic polymer, acknowledged as an excellent optical material due to its exceptional properties. High transparency in a wide range of wavelengths (from ultraviolet to near infrared) [59], excellent thermal stability and easily tailorable electrical characteristics [60], all of these augmented by low cost and wide availability, have promoted PMMA as an exceptional material used for optoelectronic devices (such as polymeric waveguides) [61], for direct-write e-beams and different micro-lithographic processes [62], and for many other optical equipment parts (e.g., lenses) [63].

In this work, we exploit tip-enhancement effects, which have mainly been used to date for imaging and for nanolithography applications (as discussed under the previous paragraph), to implement a method capable to modify the refractive index of PMMA. Considering changes in the PMMA structure that can be theoretically predicted, we hypothesize that the refractive index modifications occur as a result of localized heating and depolymerization, similar to previous efforts discussing the modulation of the refractive index of PMMA by thermal methods [64,65].

The presented results demonstrate the reported method as a viable solution for achieving local refractive index changes for sample regions sized <200 nm, with nanoscale lateral precision enabled by the use of XY sample scanning with a piezo-ceramic nano-positioning stage. While our method is demonstrated for PMMA, we argue that it can be easily tailored to address many other optical materials whose composition is prone to modifications upon laser beam irradiation.

**Materials and methods**

To extract the local refractive index changes from s-SNOM imaging data (at the interest wavelength of 1550 nm) it is necessary to use as a reference the refractive index of the unmodified PMMA sample at 1550 nm. This is obtained by employing Lorentz-Lorenz model and Sellmeier equation together with data available from the literature. Infrared absorption investigations were performed in the 930-1090 $cm^{-1}$ range to find an absorption peak necessary for improvement of the Sellmeier equation with an additional expansion term (obtaining a two-term Sellmeier equation). This improvement diminished the incongruity between data available from the literature and the original one-term Sellmeier equation. The refractive index of the bulk PMMA sample was also measured at 640 nm using the well-established Chaulnes method to confirm the accuracy of the two-term Sellmeier equation, which was further on used to obtain the refractive index of PMMA at 1550 nm. Additionally, a two-term Sellmeier equation (adjusted with the help of the Lorentz-Lorenz model and the topographical changes) confirmed the depolymerization hypothesis by predicting similar refractive index changes obtained by s-SNOM investigations. Moreover, Young's modulus measurements confirmed the material changes in terms of elasticity and volume expansion, which is expected in the case of depolymerization. All the above procedures are described in detail below.

*Sample and characterization methods*

To demonstrate the local modification of the refractive index of PMMA, with nanoscale lateral precision, we used a commercial-grade clear Plexiglas sheet of 3 mm thickness. All reported results were obtained on this proof-of-concept sample.

The characteristic parameters of the PMMA sample (necessary for refractive index measurements) are either measured or estimated from other data sources. Besides refractive index mapping at 1550 nm via s-SNOM investigations, the measured parameters are Young's modulus, refractive index at 640 nm and infrared absorption between 930-1090 cm$^{-1}$. Unless otherwise specified, the measured values are reported using mean value and standard error of the mean.

For measuring the Young's modulus, we used AFM force-distance (FD) investigations. The FD investigations were accomplished by using a Park NX10 equipment (Park Systems). The FD curves were analyzed using the XEI Data Processing and Analysis software (Park Systems) to obtain the Young's modulus by means of the Hertzian model [66]. The parameters for this model were set after running an AFM tip characterization procedure on a test structure (PA01, Mikromasch) together with blind estimation analysis in Gwyddion image processing software [67].

The refractive index of the PMMA sample in the visible domain (at 640 nm) was measured using the Chaulnes method [68] implemented in a confocal laser scanning microscope (C2+, Nikon, Japan).

The infrared absorption was measured using an s-SNOM system (Nea-SNOM, Neaspec) equipped with an $CO_2$ tunable laser illumination unit (L4GS FCCL, Access Laser) and an appropriate detection unit (HgCdTe detector, Kolmar Technologies). The implemented method for absorption is similar to the one described in [69], in which the spectra of the sample (PMMA) and of the reference material (Si, which is needed for normalization) are obtained successively. Moreover, the PMMA spectrum was normalized to the input power of each emission wavelength.

The s-SNOM investigations for refractive index mapping were performed at 1550 nm, the usual optical telecommunications wavelength. The refractive index of PMMA sample at 1550 nm was estimated at $n_{1550} = 1.4765$ using data available at [70] fitted by a two-term form of the Sellmeier equation, which is discussed later. This value was necessary for mapping the refractive index from the s-SNOM images, as described in [42,43].

*Optical setup for refractive index modification*

The proposed method for refractive index modification in PMMA relies on a low power pulsed fs laser beam that is focused on the tip of an AFM metal-coated probe in an epi-configuration, where the beam travels through the (transparent) polymer sample of interest on its way to the tip. In the absence of the AFM probe the polymer displays no modifications upon exposure to the pulsed laser beam. Conversely, when the probe is brought in close proximity of the sample (the typical range of tip-sample distances used for AFM measurements) and the laser beam is focused on its apex, laser power is enhanced in a spatial region whose extent depends on the tip's size, and composition. This enhancement results in locally confined modifications in the structure of the PMMA sample, which we found to be accompanied by refractive index changes at the surface of the sample. These refractive index modifications accomplished with our approach were quantitatively assessed with s-SNOM, according to a method that we discussed previously [43,71]. The simplified scheme of the setup is represented in Figure 1.

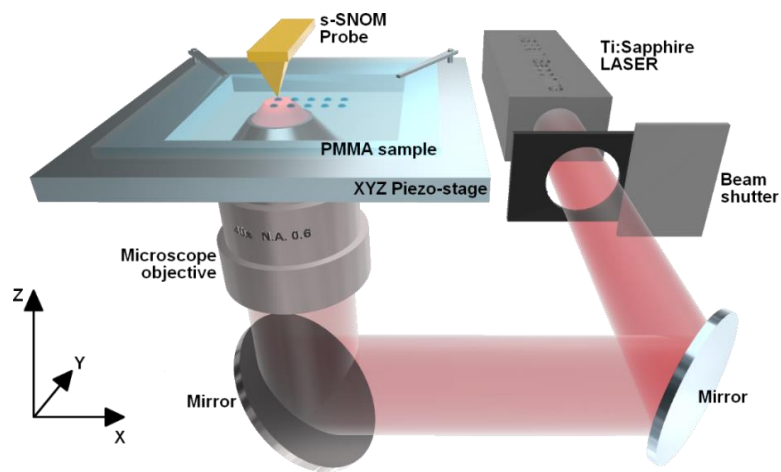

*Figure 1. Schematic representation of the optical setup for refractive index modification.*

The system used for accomplishing refractive index changes based on tip-enhanced fs laser illumination is a home-made multi-modal platform for correlative imaging with near-field and far-field modalities[72]. This imaging platform relies on an inverted Confocal Laser Scanning Microscope (CLSM, Nikon C2 confocal), featuring an Atomic Force Microscope (AFM, Q-Scope 350), which is placed on top, same as in other systems reported for correlative AFM-light microscopy applications [73]. The use of the AFM is further extended by means of a home-made module for near-field imaging. The AFM probe's position can be adjusted to match the vertical optical axis of the confocal microscope, which allows positioning the focal spot of the microscope objective on the tip apex. This imaging platform uses an XYZ piezoelectric stage (Nano LPQ, Mad City Labs) for positioning the sample with nanoscale precision on the XYZ axes, which features an orifice that allows optical access to the sample with the inverted microscope's objective. This configuration allows thus optical and scanning probe access to the surface area of transparent samples, which can hence be characterized with complementary imaging modes, working at different scales, and relying on distinct contrast mechanisms. To benefit from the tip-enhancing effect, we used a typical metal tip (Pt-coated, HQ:NSC18/Pt, Mikromasch). The input power of the pulsed laser was adjusted using a half-wave plate and a polarizer and measured at the exit of the microscope objective using a thermal sensor. For tip-illumination and refractive index patterning, we used a Ti:Sapphire fs pulsed laser (Chameleon Vision II, Coherent), with 140 fs pulse width and 80 MHz repetition rate.

*Protocol for refractive index changes based on tip-enhanced femtosecond laser illumination*

To locally modify the refractive index of the PMMA sample, we tuned the pulsed laser to 790 nm (emission peak for Ti:Sapphire laser) and in an initial step low beam power (<15 mW) was used to adjust the position of the beam on the sample surface region of interest. The light was focused on the sample surface using a 40x, 0.6 NA objective (Nikon, S Plan Fluor ELWD). Afterwards, the AFM probe was brought in close proximity of the sample's surface region where the beam was focused on by instructing the AFM probe to approach the sample for a typical contact mode AFM session. Different tip-enhancement regimes were explored by adjusting the beam power.

*Assessment of refractive index modifications*

To assess the refractive index changes occurring in the PMMA after tip-enhanced fs laser beam irradiation, we used a scattering Scanning Near-field Optical Microscope (Nea-SNOM, Neaspec), working in a pseudo-heterodyne detection configuration [74]. Imaging was performed using a 1550 nm continuous-wave laser sources (DFB Pro laser diode, Toptica). Topography, optical amplitude, and optical phase images were acquired for up to the 5-th harmonic of the tapping tip. The amplitude and phase images used for the refractive index calculations presented in this article were those collected for the 3$^{rd}$ harmonic of the oscillation frequency of the s-SNOM probe (Mikromasch, HQ:NSC16/Cr-Au). The importance of s-SNOM signal detection at harmonics of the tapping frequency has been discussed in past works [75].

The s-SNOM output images were converted into complex refractive index data (refractive index *n*, and extinction coefficient, *k*) using an algorithm based on the Oscillating Point-Dipole Model [76] (presented in our past works [43,71]). Several other methods for extracting the dielectric function and optical constants from s-SNOM images and spectra, have been meanwhile reported [41,77,78].

*Estimation of the refractive index change*

For comparing the theoretical prediction with experimental results, the refractive indices of both PMMA sample and of the sample areas locally modified due to tip-enhanced femtosecond laser exposure were estimated using the Lorentz-Lorenz (LL) model [32,79] and the Sellmeier equation [80–82]. According to LL model, the refractive index of a transparent material is closely related to the molar volume ($V_m$) and the molar refraction (*R*) of the material:

$$n^2 = \frac{1 + 2R(\lambda)/V_m}{1 - R(\lambda)/V_m}$$

The molar refraction $R(\lambda)$ is wavelength-dependent and is the sum of all atomic, bound and group contributions of the molecule unit [32]. For the sodium D line doublet (589 nm) a comprehensive list of *R(589)* values for different atomic, bound and group contributions can be found in [83]. For PMMA, the molecular unit is composed of 5 carbon atoms ($R_C$=2.418 cm$^3$/mol), 8 hydrogen atoms ($R_H$=1.1 cm$^3$/mol), 1 single bonded oxygen ($R_{O\text{-}}$=1.643 cm$^3$/mol) and 1 double bonded oxygen ($R_{O=}$=2.211 cm$^3$/mol), which totals $R_{PMMA}(589) = 24.744$ cm$^3$/mol. The molar volume for PMMA is $V_m$=85.2 cm$^3$/mol [84]. Thus, the refractive index of PMMA at 589 nm is found to be $n_{PMMA}(589) = 1.492$, which is in very good agreement to experimental measurements [70].

The molar volume for the depolymerized material was estimated by evaluating the volume expansion (calculated from topography data) proportionally to the same number of moles. First, it was necessary to estimate the sample depth at which the depolymerization occurs. This depth should be related to the s-SNOM imaging depth because the depolymerization occurs only at the proximity of the s-SNOM tip due to field enhancement. In the literature it is reported that the imaging depth in the case of s-SNOM can be of tens of nanometers [85–87]. In our case, an estimated depth of 60 nm (similar to reported depth for water [87]) fits best the calculated refractive index to the measured refractive index via s-SNOM. Additionally, the molar volume was adjusted accordingly to the height of each modified sample area, which depends on the laser pulse energy.

When local depolymerization occurs, the molar volume increases (as expected) and an additional C=C bound appears causing an increase of R with $R_{C=C}=1.733$ cm$^3$/mol [32], resulting a total $R_{MMA}(589)$ of 26.477 cm$^3$/mol. With a molar volume of 106 cm$^3$/mol [84], the refractive index of the depolymerized material is $n_{MMA}(589)=1.414$.

To determine the molar refraction at 1550 nm (at which the refractive index was experimentally obtained based on s-SNOM imaging) we used the Sellmeier equation [80–82] to fit the discrete data available at [70]. Generally, the Sellmeier equation is written as:

$$n^2(\lambda) = 1 + \sum_i \frac{A_i \lambda^2}{\lambda^2 - l_i^2}$$

Here, $A_i$ stands for strength coefficient of absorption resonance at wavelength $l$ expressed in micrometers. Usually, one or two terms in the sum are sufficient to accurately estimate the refractive index far from absorption resonances. For PMMA, based on data from [88], M. N. Polyanskiy obtained a one-term equation [89]. To enhance the accuracy, we added a second term in the sum specific to the mid-infrared absorption line. The infrared absorption investigations that we carried out revealed an absorption peak at 1064 cm$^{-1}$ (see Results), and the corresponding supplemental term was added in the equation. After fine adjustments of the coefficients $A_i$ to fit the data from [70], the improved Sellmeier equation for PMMA is:

$$n^2(\lambda) = 1 + \frac{1.1897\lambda^2}{\lambda^2 - 0.011313} + \frac{0.55\lambda^2}{\lambda^2 - 88.36}$$

Connecting this equation to the LL equation for $\lambda = 1550$ nm, the molar refraction of PMMA becomes $R_{PMMA}(1550) = 24.25$ cm$^3$/mol.

For the depolymerized areas, we consider a similar equation of $n$ with adjusted strength coefficients to fit the refractive index value at 589 nm. With this amendment, the molar refraction at 1550 nm becomes $R_{MMA}(1550) = 25.98$ cm$^3$/mol, which is further on employed into LL model and Sellmeier equation to predict the refractive index changes for the case of depolymerization and to compare the predicted results to those obtained by s-SNOM measurements.

**Results**

*Infrared absorption and refractive index*

The mid-infrared absorption data acquired on the PMMA sample in the range of 930-1090 cm$^{-1}$ is represented in Figure 2 (a). The black dots represent the experimental values, while the dashed line represents the resulting spline interpolation of these data points, and the solid line depicts previously reported data [70]. The plots are normalized for comparison. Two peaks emerged from the investigations, both specific to PMMA material: one around 980 cm$^{-1}$ and another one around 1064 cm$^{-1}$. According to data available in the literature [90,91], the two peaks correspond mainly to the C–O stretching and in-plane bending of OCH$_3$ (for band around 980 cm$^{-1}$) and to C–C stretching (for band around 1064 cm$^{-1}$).

The absorption peak at 1064 cm$^{-1}$ (which is better defined than the other) was used to supplement the Sellmeier equation with an additional term, as exposed in the Methods section. In Figure 2 (b) we can observe how the new equation improves the data fitting. The discrete data from reference [70] is represented with red starry dots, the one-term Sellmeier variation with blue dashed line and the two-term Sellmeier variation is represented with black dashed line. The improved Sellmeier equation allowed us to find and use a more precise refractive index value at 1550 nm which correctly fits the available discrete data. In the same figure are represented the measurements of PMMA refractive index at 640 nm (considering the mean and standard deviation), which are in are good agreement with the two-term Sellmeier plot ($n_{PMMA}(640) = 1.4904 \pm 0.0004$).

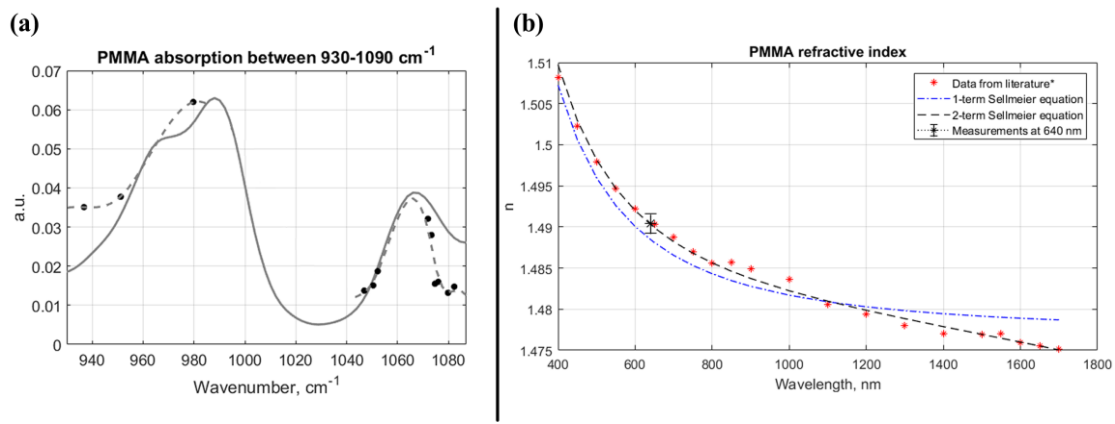

*Figure 2. (a) Infrared absorption of PMMA between 930-1090 cm$^{-1}$: solid line results from data available at reference [70], black dots represent the experimental absorption values and the dashed line is the spline interpolation of the experimental data. (b) PMMA refractive index variation with wavelength. The one-term Sellmeier variation is represented with blue dashed line and the two-term Sellmeier variation is represented with black dashed line. The refractive index measurement at 640 nm is represented in the figure by starry mark with error bar.*
*\*the discrete data from reference [70] is represented with red starry dots.*

For the investigated PMMA sample we measured a Young's modulus of $2.61 \pm 0.3\ GPa$ (which is in good agreement to other reported Young's modulus values of PMMA [92]).

For the sample regions exposed to pulsed laser, the Young's modulus was $0.21 \pm 0.03\ GPa$. This result represents a substantial change of the material in terms of elasticity. The local decrease of the Young's modulus can be linked to volume expansion [93], which could explain the observed topographical local height increase of the corresponding sample regions. We find worthy to mention that volume expansion is expected in case of depolymerization [32].

*Complex refractive index modifications*

The structural changes of the PMMA sample were assessed at six different laser pulse energies of 0.375 nJ, 0.400 nJ, 0.425 nJ, 0.450 nJ, 0.475 nJ and 0.500 nJ, nine sample regions being modified for each pulse energy. For each illumination condition (and each corresponding sample region), a 5 seconds laser exposure interval was used. Figure 3 (a)-(c) presents the s-SNOM (3$^{rd}$ harmonic amplitude and phase) and AFM images acquired on the modified sample regions. In the visible array of modified sample regions, each row of modified domains was created with the laser pulse energy indicated at the left side of the Figure 3 (a).

Each pair of s-SNOM amplitude and phase images cropped around the modified domains was collectively post-processed to extract the refractive index and extinction coefficient maps, using a previous described protocol [43,71] mentioned in the Methods section. The refractive index analysis showed that the six sample regions exhibit different values of the complex refractive index (see Figure 3 (d) for refractive index and (e) for extinction coefficient).

At the locations where the sample was exposed to the pulsed laser in the proximity of the sharp tip, the refractive index decreased substantially, with some values close to 1.30. On the other hand, the extinction coefficient increased, though it remained in the same order of magnitude.

Given that the AFM and s-SNOM images are by default registered, as s-SNOM and AFM investigations are performed simultaneously, we used the topography image to segment each modified PMMA domain in both the refractive index and extinction coefficient maps. The use of the topography image as reference for segmentation was required in order to evaluate the same areas in terms of refractive index and extinction coefficient modifications. Thus, the area of each modified domain is defined based on the local maximum in topography image and the surrounding area with elevation higher than 10% of the local maximum. Figure 3 (f) shows a single modified PMMA domain that we fabricated at 0.375 nJ laser pulse energy. It illustrates how the locations of the topographical and optical changes overlap, with the white curve marking the boundary of the modified domain. Interestingly, the modified extinction coefficient covers an area smaller than the topography domain, the explanation for this result probably being linked to the gaussian profile distribution of the pulsed laser beam. The size of the domain measured in the topography image is less than 200 nm which is about four times smaller than the wavelength of the laser beam used to create it.

Following the morphology assessment, we calculated the mean refractive index and the mean extinction coefficient for each modified PMMA domain. Similar to the topography features analysis, the average refractive indices and extinction coefficient were calculated for all modified domains at the same laser pulse energy. These results are compared to the refractive index (and the extinction coefficient) corresponding to the areas not exposed to the pulsed laser beam located around each modified domain, with the same area in terms of number of pixels. Variations of refractive index and extinction coefficient with the laser pulse energy are represented in Figure 3 (g).

The refractive index change estimated using the LL model and considering different levels of volume expansion due to depolymerization is represented as well in Figure 3 (g) with starry green dots. The results are represented as variation with different laser pulse energies.

*Topography and elasticity*

Using the proposed protocol for tip-enhanced illumination no modifications of the material have been observed to occur for beam power levels lower than 15 mW. The gradual increase of the laser pulse energy resulted in an enhancement of the electric field at the tip of the probe, which in turn resulted in structural modifications of the sample surface underneath, which were accompanied by refractive index changes. Tip-enhanced illumination resulted also in small topographical changes which were observed with the AFM system, by visualizing the AFM setpoint level.

For each modified domain the maximum height was extracted together with its diameter using topographical measurements in the AFM image. We find it noteworthy to highlight that all modified domains exhibit higher height compared to the unilluminated sample surface and nearly circular shapes in the horizontal plane, Figure 3 (c).

The impact on the topography was evaluated by measuring the average maximum height and diameter for all PMMA domains modified with the same laser pulse energy. The results are presented in Figure 3 (i) and (j), showing an increase of both height and diameter with the laser pulse energy.

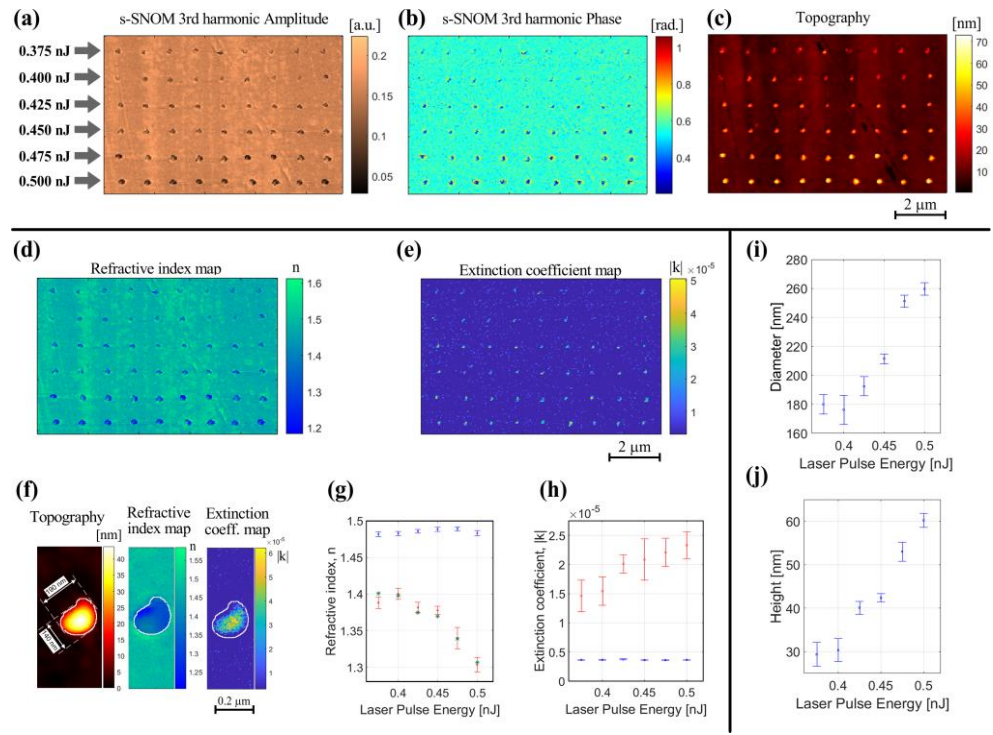

*Figure 3. Surface investigations on the PMMA modified regions by s-SNOM and AFM. (a) 3$^{rd}$ harmonic amplitude and (b) 3$^{rd}$ harmonic phase images acquired by s-SNOM investigations; (c) topography image acquired by AFM; (d) refractive index n map; (e) extinction coefficient k map; (f) confirmation of the overlapping between optical and topographical surface changes; (g) refractive index n of the modified domains (red) compared to unmodified areas (blue), for six different laser pulse energies. The predicted refractive index change based on LL model is represented with green starry dots; (h) extinction coefficient k of the modified domains (red) compared to unmodified areas (blue), for six different laser pulse energies; (i) variation of topographical diameter of the modified domains as function of laser pulse energy; (j) variation of topographical height of the modified domains as function of laser pulse energy.*

## Discussion

Previous works reported the successful refractive index change in PMMA via thermic processes or laser ablation [22–24,26–29]. The refractive index modification inscribed on the surface of the polymer sample due to exposure to fs laser in the presence of a nano-sized tip occurs at very low pulse energies. The effect is possible because of the tip enhancement effect, which can be responsible of an enhancement factor of several orders of magnitude [94,95]. This results in highly increased optical power because of the dependence of the power to the squared electric field intensity, hence the occurrence of the refractive index modification in low power regime in the presence of a metallic tip. In terms of the laser beam power levels (or laser pulse energy) used to induce material changes, the range was tens to hundreds of nJ [24,28]. By exploiting valuable tip-enhancement effects, the proposed method allows the use of laser pulse energy that are two orders of magnitude lower. Moreover, to the best of our knowledge, in all previous studied, the size of the modified domains was always larger than the wavelength of the laser beam, ranging from 1 micrometer to several tens of micrometers in diameter [9,22,23]. In this work, we report modified refractive index domains with < 200 nm diameters at 790 nm illumination (~4 times lower than the illumination wavelength).

We report negative changes of the refractive index, which may be also correlated to the MHz repetition rate regime of the pulsed laser [59]. Negative changes are reported for MHz repetition rates and positive changes of the refractive index for kHz repetition rates [59].

From the curves plotted in Figures 3 (g)-(j) one can observe that local modification of the refractive index is accompanied by topographical modifications of the sample's surface. This effect is reported in other published works without tip enhancement [30,96], which suggest that the topography modifications are influenced mostly by the pulsed laser. However, the impact of the pulsed laser on the topography of the sample is minimal: tens of nanometers of elevation and about 200 nm or less in diameter for lower energies (see Figures 3 (i) and (j)). The areas of the modified domains are less than $\lambda/4$ in diameter for low energies, in spite of low numerical aperture (0.6). These results are very promising for optical materials and nano-sized optical devices, where nanoscale refractive index changes are required.

In the presented experiment, the refractive index decreases with the increase of the laser pulse energy: increasing the laser pulse energy from 0.375 nJ to 0.500 nJ, the mean refractive index decreases from 1.389 to 1.301. Regarding the extinction coefficient, its absolute value increases from $0.37 \times 10^{-5}$ to around $2 \times 10^{-5}$, having a small increase with laser pulse energy. This can have an impact on the light absorption in the material, which can be overcome easily using micro- and nanoscale devices (due to small distances). Nevertheless, this effect has no impact in developing bi-dimensional photonic crystal lattices (PCL) and PCL-based waveguides. An important observation based on Figure 3 (f) is that the area with modified extinction coefficient is smaller than the topographical modified area or the modified refractive index area. While this effect may be connected to the gaussian profile of the beam, further investigations will be needed to better understand the effect of tip-enhanced pulsed laser beams over the extinction coefficient of PMMA.

The depolymerization effect under exposure to pulsed laser beams was reported as well in other studies [22,23]. Under this assumption, we employed the LL model and the Sellmeier equation to estimate the refractive index of modified areas and we found a good agreement with the s-SNOM measurements. This result enforces the idea that the depolymerization effect occurs under tip-enhanced pulsed laser beam condition.

**Conclusions**

In this work we present a method based on tip-enhanced fs laser illumination to modify the complex refractive index for nano-sized domains in a PMMA sample. While previous related methods reported to date were not able to induce modifications for domains sized below the illumination wavelength, we report circular domains with diameter <200 nm for excitation at 790 nm. This four-fold increase in precision holds significant potential to pave the way for a wide palette of novel applications that can benefit of optically induced super-resolved refractive index modifications. The refractive index of the modified domains was evaluated with nanoscale lateral precision using a s-SNOM system operating at 1550 nm (a common wavelength in telecommunications industry) and was found to be lower compared to the refractive index of the control sample, and the extinction coefficient was found to be about four-fold larger. While the extinction coefficient seems to be less dependent to the laser pulse energy, the refractive index was found to decrease considerably with the increase of the laser pulse energy. While previous methods used to modify the refractive index of PMMA report the use of pulse energy levels in the range of tens to hundreds of nJ, the here proposed approach allowed the use of laser pulse energies that were two orders of magnitude lower. Given the achieved performance in terms of size of domains with modified refractive index and pulse energy requirements, we consider the developed method to be highly relevant for nano-modification and nano-fabrication of optical devices, surface light waveguides, bidimensional photonic crystals devices, and for nano-photonics in general.

**Funding**

This work was partially supported by the UEFISCDI grant PN-III-P2-2.1-PED-2019-2386 (INTEGRAOPTIC). The use of the Neaspec NeaSNOM Microscope and of the Chameleon Vision II (Coherent) Ti:Sapphire laser was possible due to European Regional Development Fund through Competitiveness Operational Program 2014–2020, Priority axis 1, Project No. P_36_611, MySMIS code 107066 - INOVABIOMED.